# Roadmap to a Roadmap: How Could We Tell When AGI is a 'Manhattan Project' Away?

**John-Clark Levin**[1] and **Matthijs M. Maas**[2]

**Abstract.** This paper argues that at a certain point in research toward AGI, the problem may become well-enough theorized that a clear roadmap exists for achieving it, such that a Manhattan Project-like effort could greatly shorten the time to completion. If state actors perceive that this threshold has been crossed, their incentives around openness and international cooperation may shift rather suddenly, with serious implications for AI risks and the stability of international AI governance regimes. The paper characterizes how such a 'runway' period would be qualitatively different from preceding stages of AI research, and accordingly proposes a research program aimed at assessing how close the field of AI is to such a threshold—that is, it calls for the formulation of a 'roadmap to the roadmap.'

## 1  RUNWAY TOWARD AGI

A large proportion of AI researchers expect that artificial intelligence capable of outperforming humans in many or all tasks will be within reach within four or five decades [24,48]. While the literature at times makes slight distinctions between 'high-level machine intelligence' [24], 'human-level AI' [45], or 'strong AI' [39], the most common term for capturing most of these concepts is 'artificial general intelligence' (AGI) [21]. In essence, AGI denotes broad and flexible AI systems that are at least at human level in every cognitive domain [4].

AGI is closely related to the concept of 'superintelligence'—AI that greatly surpasses the best humans at every cognitive task [4]—for two reasons. First, a human-level AI would also at least have all the advantages computers already have. For example, thanks to fast data processing and vast memory, an AI capable of comprehending even high school-level English could memorize all the knowledge on Wikipedia with trivial ease. Thus, AGI would inherently be superhuman in many or most areas. Second, it is often projected that AGI could quickly become superintelligent through a process of recursive self-improvement sometimes known as an 'intelligence explosion' [4,15,22,40]. For these reasons, the goal of achieving AGI entails far greater potential capabilities than "mere" human-level intelligence.

Already today, several dozen R&D projects worldwide are actively pursuing AGI [3]. Scholars have observed that the enormous benefits of human-level AI, possibly including "radically transformative" [25] impact would be comparable to the Industrial Revolution [36]. The economic, scientific, and military advantages conferred by such a breakthrough, along with the prospect of unlocking an intelligence explosion, could mean geopolitical dominance for the first nation state with access to AGI. This possibility may motivate an international "arms race" toward this technology [2,4,27]. That potential dynamic has serious implications from the perspective of technical safety [1,17,59] as well as in the context of governance and policy [11,13,50,68].

These strategic considerations have fostered analogies to the development of nuclear weapons [42,55]. Some observers have compared the state of AI research to nuclear physics in the 1930s [23], with the suggestion that a massive Manhattan Project-style engineering effort could allow one well-resourced actor to achieve AGI much sooner than rivals expect.

Critics of this comparison have correctly highlighted differences between the open culture of contemporary AI research and the coordinated and covert nature of wartime nuclear weapons development [55]. Further, there is good reason to believe that the state of basic research on AGI is currently much more immature than the state of nuclear physics just prior to the Manhattan Project [26]. But these disanalogies need not remain true forever.

As the scientific roadmap becomes clearer, as it did for nuclear weapons by 1942, a 'runway' toward AGI may come into focus—where the key theoretical breakthroughs have been made, and the remaining research questions are well enough formulated to bring human-level AI within reach of a massively funded engineering project. While this is not a certainty, the plausibility of such conditions occurring calls for better understanding of how such a runway might be detected and what its implications could be.

### 1.1  Defining a 'runway sprint' project

In considering the scale of resources that might be applied once a runway to AGI appears, two historical examples stand out: the Manhattan Project and the Apollo program. In both cases, a powerful state actor perceived a technological opportunity of paramount strategic importance, and at peak devoted about 0.4% of GDP toward greatly accelerating progress toward it. Scaled as a percentage of current U.S. GDP, that amounts to something like an annual budget of $80 billion in today's dollars [61].

These examples stand in stark contrast to other scientific megaprojects undertaken when the strategic stakes are much lower. The Laser Interferometer Gravitational-Wave Observatory (LIGO) cost less than $33 million a year over 19 years of development [6]. The Human Genome Project took 13 years and spent the equivalent of $400 million a year in today's dollars [49]. The Large Hadron Collider (LHC) at CERN took about 10 years to build, at an average cost of about $475 million per year [7,38]—though full operational costs are in excess of $1 billion per year [56]. Finally, the ITER fusion reactor is projected to average just under $2 billion per year

[1] Department of Politics and International Studies, University of Cambridge, United Kingdom. Email: johnclark.levin@gmail.com

[2] Centre for International Law and Governance, Faculty of Law, University of Copenhagen, Denmark. Email: Matthijs.Maas@jur.ku.dk

during 12 years of construction [18].[3] Moreover, rather than single-country projects, the costs of these programs have each been shared among several countries.

While these efforts are arguably comparable to the atomic bomb or the moon landing from a purely engineering standpoint, none of them had similar geopolitical urgency. And although LIGO, the LHC, and ITER all received substantial international public funding, they were not fundamentally outside the capacity of non-state actors—costing much less than the largest private-sector R&D budgets, such as those of Amazon and Alphabet [14].

So from the limited sample of historical precedents, it appears plausible that AGI—if its implications are properly understood by policymakers—could motivate investment more similar to the quest for nuclear weapons than to the effort to detect gravitational waves.[4] Indeed, current military spending evinces governments' willingness to pour enormous resources into projects when there is a perceived strategic imperative. The F-35 stealth fighter program, for example, is projected to cost the United States around $16 billion annually over its 60-year lifespan [20].[5] This aircraft, while an evolutionary improvement over its predecessors, offers nowhere near the revolutionary defense advantages that AGI would bring to numerous fields from cyberwarfare to autonomous weapons [19,54,55].

There is no bright-line definition of what would constitute a Manhattan Project-level AGI effort, but very roughly we might consider this a program with resources 1-2 orders of magnitude greater than existing projects. This would be effectively beyond the means of the private sector, and only be plausible for several of the world's wealthiest state actors. Such a project would be intended to 'sprint' down an apparent runway to artificial general intelligence within a relatively small number of years,[6] likely under perceived strategic pressure. While the Manhattan Project involved physical sequestration of top scientists at a remote government research facility, an AGI sprint project need not take this particular form. It could involve clandestine support of private sector actors, and because computing does not require as much concentrated infrastructure as do physics experiments, its activities could be highly distributed geographically. Rather, the essential similarities to the Manhattan Project would be massive funding and strategic urgency.

## 1.2   Runway sprints and surface area

Whether a runway sprint succeeds likely depends partly on when it is initiated. The U.S. government could have devoted not 0.4% but 4.0% or 40% of its GDP to exploring "military applications of nuclear physics" in 1932, yet probably not have achieved a working bomb by 1935. The key reason is that the basic science was still too immature, and the problem had not yet been broken down into a roadmapped series of sub-problems (e.g. enrichment, isotope separation, metallurgy, explosive lenses [30]) to which engineering effort could be applied. Regardless of a state actor's willingness to commit resources, it was simply not clear in 1932 how to allocate them.

This highlights the more general idea that problems in science and technology have a 'surface area' [12], which characterizes how readily divisible they are for researchers and resources to bring simultaneous efforts to bear in solving them. Recalling the truism that 9 women cannot gestate a baby in 1 month, problems with small surface area may require a sequence of abstract theoretical breakthroughs by the very top thinkers in a field. Simply throwing large numbers of PhDs at the problem likely will not overcome those causal dependencies and allow a quick breakthrough.

By contrast, problems with larger surface area have well-defined and roadmapped sub-problems that can be worked on in parallel. Under those conditions, greatly increasing the resources devoted to the problem can rapidly accelerate progress. In other words, as surface area expands, it is easier to "buy progress" toward completion. Conceptually, then, AGI research would be on the runway and thus susceptible to a sprint project when advances in theory and basic research expand the surface area of the problem enough that a roughly Manhattan Project-size injection of resources could greatly accelerate the timeline and achieve success within several years.

## 1.3   Surface area of AGI

The surface area of the AGI problem can be understood as evolving along a continuum, from smallest to largest. In its initial state, before Alan Turing's seminal 1950 paper "Computing Machinery and Intelligence," [63], the field was essentially waiting for someone to come along and conceptualize artificial intelligence itself. At that stage, resource application was probably almost irrelevant to progress, as the problem itself wasn't formulated yet. At the other end of the continuum is some hypothetical future point where all the scientific questions have been answered and success is *purely* a matter of resource application—perhaps physically constructing a certain piece of hardware, feeding in data, or even re-assigning sufficient pre-existing supercomputer or cloud computing capacity.

As of 2020, it is unclear where the field stands between those two poles. It could be that there are many profound theoretical problems still to be solved [33,43]—perhaps some that we haven't even dimly imagined yet. It is also possible that the surface area has already expanded enough that a sprint project begun now would likely succeed. While it isn't possible to precisely measure the scientific surface area of AGI, as we argue in section 3.1, further research may be able to estimate when the runway has been reached.

---

[3] Even the most expensive single object in history, the International Space Station, cost about $11-15 billion per year over 10 years [16]—well below the GDP-scaled costs of the Manhattan Project and Apollo program. This is unsurprising, since the ISS had much lower strategic importance.
[4] There have been some abortive attempts at moonshot-style AI megaprojects before. DARPA's 1983-1993 Strategic Computing Initiative in the United States spent about $100 million a year in pursuit of what we would now call AGI [58]. But the SCI was undertaken with a very primitive understanding of what AGI would entail, and DARPA effectively misjudged that it was on the runway. Something similar may well happen in the future. But the fact that states may incorrectly believe themselves to be on the AGI runway does not imply that eventually one of them won't be correct. Even if 10 runway sprints are ultimately attempted, 9 are premature and only 1 succeeds, this scenario will have been worth taking seriously.

Indeed, this possibility of misjudgment underscores the need for better means of understanding the dynamics of the runway and assessing the field's proximity to AGI [11].
[5] For comparison, perhaps the most startling AI advance over the last year has been GPT-3, OpenAI's powerful text-generating language model [5]. Training GPT-3 cost an estimated $12 million in compute [64,66]. That's less than the cost of the missiles carried by one fully loaded F-35 and the helmet the pilot uses to aim them [46,47,65].
[6] The meaning of "small number of years" is necessarily somewhat vague here, but might be loosely defined as shortening a decade-plus trajectory to within something like 3.5 years (Manhattan Project) or at most 8 years (Apollo program).

## 1.4 Three runway scenarios

In assessing progress toward AGI with respect to the runway concept, we note three broad possibilities. In short: 1) there will be a runway period, but we are not in it yet; 2) there is a runway period, and we are in it already, or; 3) there will be no runway period, because the "last mile" of the AGI problem is not susceptible to large-scale engineering. After sketching each in further detail, we will argue why the first two likely occupy enough of the probability space to make further research into AGI runway conditions warranted.

### 1.4.1 We're approaching the runway

The first possibility, and the one we consider most likely given the evidence, is that there will eventually come a moment when the surface area of the AGI problem has expanded enough to allow a successful runway sprint project, but that more basic research needs to happen first before we approach such a point. In this scenario, the impressive AI progress of the past decade does not imply that general intelligence can be attained by just scaling current techniques up. Rather, there might remain 'computational primitives' fundamental to human-level cognition that are yet to be discovered and theorized [44], or essential aspects of intelligence itself that remain deeply misunderstood. That does not mean that steady progress over the coming decades may not crack these problems. It simply means that no party today has the sort of engineerable roadmap to AGI that would constitute a runway.

### 1.4.2 We're already on the runway now

The second possibility is that the relevant theoretical breakthroughs for AGI have already been made, even if this is not widely apparent or recognized yet. For example, it could be that Rich Sutton's "Bitter Lesson" [62] is true, and that the computationally intensive "scaling" approach notably advanced by OpenAI will prove sufficient to achieve AGI [29,35]. According to this hypothesis, the main factor preventing a system like GPT-3 from achieving human-level performance is simply that it does not yet have enough parameters. If so, then a reasonably straightforward process of adding orders of magnitude more parameters, training data, and compute[7] would result in AGI—whether this would come at the level of GPT-5 or GPT-12.[8] This would imply that we are already on the AGI runway as of 2020, which would suggest two main possibilities. The first is that state actors like the United States and China *could* achieve AGI substantially sooner than OpenAI itself can by applying their resources aggressively, but they have not recognized this opportunity. The second is that one or more such actors have recognized this opportunity and already commenced runway sprint efforts in secret.

### 1.4.3 There is no runway

The third possibility is that there will be very substantial overhangs in hardware and software engineering, such that AGI is solved quite suddenly by one final, keystone theoretical breakthrough. We might imagine some discovery in another discipline, or some new algorithmic technique—which is nearly costless to implement—that quickly activates latent potential in systems that have already been designed or built. In this scenario, development toward AGI proceeds at something similar to current rates, with ever-improving capabilities, but no clear roadmap to AGI ever appears. Then, following some hard-to-predict single discovery in mathematics, cognitive neuroscience, or theoretical computer science, it becomes possible to use existing technology, or recombine existing approaches, in a new way that promptly enables AGI. In a case like this, the surface area of the problem never expands enough to make Manhattan Project-scale resources particularly useful.

## 1.5 Assessing the probability space

It is of course possible that no AGI runway will ever manifest. If not, this would be impossible to determine in advance—the counterproof would be the sudden appearance of AGI itself. But such a trajectory would be quite surprising. All the most comparable precedents—Manhattan Project, Apollo program, LIGO, Human Genome Project, and LHC—were brought to fruition with large-scale engineering effort. There is no compelling reason to suspect *a priori* that AGI will be different.

If there is *some* point at which AGI will fall within reach of a runway sprint, the remaining question is whether this has happened yet. Here, the behavior of relevant actors provides a fairly strong signal. Almost none of the top AI labs and researchers in the field have described AGI as fully theorized or purported to present a full engineering roadmap. Even OpenAI, the lab most closely associated with the scaling model, has avoided making strong public claims about whether present approaches can be scaled up all the way to AGI.

Furthermore, no AI labs have secured even multibillion-dollar annual funding. This is significant because five major tech companies have R&D budgets over $10 billion a year: Amazon, Alphabet, Intel, Microsoft,[9] and Apple. So if any research group could make a convincing case that AGI was within sprinting distance, it is reasonable to expect that deep pockets in Silicon Valley would at least be investing far more heavily than they currently are.

Likewise, state actors do not appear to believe that the field has reached the AGI runway. There is no indication that anyone has already begun a sprint megaproject. Given the relatively open nature of the international AI research community [34,55], it would presumably have at least sparked persistent rumors if the U.S. or China had secretly committed to such an undertaking. Yet there seem to be no credible voices raising those suspicions.

It is certainly conceivable that all these actors—including the world's top AI experts and governments with an intense interest in achieving technological advantages over rivals—are overlooking present runway conditions. But the fact that they appear to be signaling a consensus with their choices is significant evidence about the information available to them and how they have analyzed it. The most parsimonious explanation is that this consensus holds because the surface area of the AGI problem has not yet expanded enough—that we have not yet reached the runway.

Taken together, these factors suggest a very significant chance that either the first or second scenario holds. Since, as we describe in section 2, the implications of reaching the AGI runway may be dramatic, there is good cause for a research program aimed at assessing how close such a runway may be.

---

[7] In this scenario, there would likely still be some theoretical challenges entailed in this scaling-up process. But they would be the sort of challenges that a large number of well-paid computer scientists could be confidently expected to eventually solve, rather than deeper fundamental issues whose ultimate tractability is substantially in doubt.

[8] As joked by Geoffrey Hinton in reaction to the GPT-3 announcement: "Extrapolating the spectacular performance of GPT3 into the future suggests that the answer to life, the universe and everything is just 4.398 trillion parameters" [31].

[9] In 2019, Microsoft did make a multi-year $1 billion investment in OpenAI [51].

## 2 INCENTIVES ON THE RUNWAY

As of 2020, AGI remains relatively poorly theorized. Not only have the key research questions not been solved, many of them have likely not even been formulated yet. We can speak of AGI in functional terms (what it could *do*) [21], but it remains broadly unclear what neurocomputational mechanisms are actually *behind* human general intelligence, or what engineering challenges would be entailed in emulating them digitally. With all this uncertainty, AGI has attracted only a small share of the mainstream attention now going to artificial intelligence as a looming geopolitical issue.

While state actors like China [8] and the United States [52,53] have expressed long-term strategic interest in artificial intelligence, this is largely focused on below-human-level AI applications, and there is scant indication that either fears the other might achieve strategic surprise with AGI in the foreseeable future. Meanwhile, there is a burgeoning global movement encouraging responsible development around issues like ethics, bias, transparency, and autonomous weapons [3,9].

In this climate, it is difficult for policymakers to see AGI as either an imminent threat or a decisive opportunity—and comparatively easy to support cooperative international AI governance regimes [9]. This is because the net incentives for cooperation are currently relatively good (e.g. favorable public relations, lessened competition), and the short-term rewards of defection are comparatively modest and non-decisive (e.g. better cyberwarfare tools), whereas the long-term rewards of defection, while large (AGI and "decisive strategic advantage" [4]), are far-off and uncertain. This may create a false sense of security in the AI governance community—the perception that if states can be convinced to avoid an all-out AI arms race in the near term, they will continue to hold back indefinitely.

Yet if the viability of a runway sprint project depends chiefly on a few surface-area-expanding theoretical breakthroughs instead of smoothly-scaling inputs, we might instead see sudden and discontinuous incentive shifts. Historically, the prospect of disruptive new technologies can suddenly upend the strategic calculations of states, leading to an erosion or the functional obsolescence of existing governance regimes [10,41]. Likewise, the continued stability of any AI governance equilibrium may hinge on states repeatedly asking themselves the binary question: "can we (for any amount of funding) sprint down a runway to AGI?"—and answering in the negative.

Once state actors realize that basic research has generated the roadmap of relatively engineerable sub-problems that would constitute a runway, that answer would flip from 'no' to 'yes.' When it thus appears that the geopolitically decisive advantages of AGI or superintelligence can be attained quickly by greatly scaling up investment, the classic one-off prisoner's dilemma would take hold [60]. That is, not only will states be tempted by the rewards of achieving AGI first, they might reason "if we can sprint to AGI, so can the other side"—prompting an immediate preemptive sprint, even if there remains some uncertainty around whether AI science has in fact reached the AGI runway.

Thus, whether one is already concerned about an imminent "AI arms race" [28,37], believes that this risk is overblown [57,69], or is optimistic that such a race might be controlled [32,42], the emergence of AGI runway conditions would greatly exacerbate these dynamics, and introduce new sources of instability. Crucially, the incentives state actors would face while on the runway would push them toward all-out competition even if none of them would have independently sought AGI for offensive or hegemonic purposes.

## 3 TOWARD A ROADMAP TO THE ROADMAP

If one or more state actors suddenly undertake runway sprint projects toward AGI, the incentives to rush could compromise AI safety [2], and national security policies might shift in ways (e.g. changes to immigration, intellectual property, export controls) that disrupt ongoing research in the private sector and academia. Worse, if democratic regimes ignore signs that the field has entered the runway period, an authoritarian power might secretly undertake a successful sprint and achieve global dominance. In either case, these risks might be reduced by a broadly-shared set of metrics for assessing how close we are to the runway—that is, how close we are to an AGI roadmap that would enable a sprint.

Measuring this progress could provide a 'fire alarm' for AGI [67]; it would help other stakeholders take appropriate precautions and potentially influence the incentives states face. A well-defined roadmap toward the AGI roadmap could also help democratic policymakers assess the likelihood that rival authoritarian regimes have secretly begun such projects—and could thereby motivate collaborative international efforts to prevent them from succeeding first. This consideration likely outweighs concern that such a meta-roadmap could act as an information hazard. Better for everyone to be asking these questions openly than have state actors groping privately for answers amidst high uncertainty.

### 3.1 Measuring surface area

The fundamental question is: what needs to happen before the surface area of the AGI problem is large enough that a state actor's runway sprint project could plausibly tackle the remaining steps within several years? Flowing from that: how can we assess the field's progress toward crossing that threshold? This suggests need for further focused study of 'runway' conditions within AI assessment research. While such work will no doubt refine the relevant criteria, several broad classes of signals may provide promising insight about whether the surface area has expanded enough for there to be an engineerable roadmap.[10]

#### 3.1.1 *Roadmapped sub-problems*

In order for vast resources to be able to scale up research impact, there has to be a well-articulated set of sub-problems which, if separately solved, would cumulatively result in AGI. Implicit in this is the need for a better-theorized understanding of what AGI itself would even entail: a research goal comparably clear to detonating an atomic bomb or landing a man on the moon. No roadmap will be perfect or free of uncertainty, but understanding the general path dependencies makes a large-scale engineering project plausible. Thus, as the field gains increasing clarity about the key sub-problems to AGI, it will be closer to the runway.

---

[10] These criteria are necessarily somewhat vague, because if they were fully concrete and fleshed out, it would already be clear how close we are to the AGI runway. Devising a useful "roadmap to the roadmap" will likely require ongoing research collaboration between AGI theorists, managers of previous scientific megaprojects, and the sorts of policymakers who would be charged with actually evaluating whether an AGI roadmap is robust enough to justify a massive runway sprint project. This paper is merely intended to highlight promising areas for more extensive and systematic study.

### 3.1.2 *AGI production function*

An actionable roadmap would provide clear estimates of the 'production function' for AGI: "[t]o what extent does [AI] performance scale with training time, data, compute, or other fungible assets?" [11]. That is, it would empirically formalize the relationship of resource inputs to results or breakthroughs. A well-theorized function would let a state actor make an evidence-based allocation of its engineering resources between, for example, hardware, algorithms, and data collection.

### 3.1.3 *Capital intensiveness*

Related to the above, progress toward AGI can be seen as requiring a combination of scientific talent to solve theoretical problems and the fungible resource investments characterized by the 'production function.' While resources like compute and basic hardware can be scaled up arbitrarily with the size of financial investment, human research discoveries are inherently more uncertain [67]. Hiring twice as many high-level PhD researchers does not cleanly scale to twice as many theoretical advances. Thus, to the extent that progress toward AGI depends on investments in hardware and software capital, the surface area is larger and a runway sprint project is closer to viability.

### 3.1.4 *Parallelism*

Ambitious research programs aiming at technologies like nuclear weapons or AGI will inevitably hit roadblocks. Problems not apparent at the outset will impede progress, and new approaches will be needed. Therefore, when there are several apparent parallel approaches to potential bottlenecks, the surface area is greater than when progress is highly dependent on one single speculative advance. This makes progress more robust to individual failures, and when more candidate approaches that can be attempted in parallel, large-scale resources can be brought to bear more efficiently. Roughly speaking, this translates to creating multiple research units to work on different sub-problems at the same time. By contrast, when known sub-problems must be tackled sequentially due to path dependencies, it is generally harder for investment to accelerate progress.

### 3.1.5 *Feedback speed*

The goal of a sprint project is to trade money for time—using enormous resources to accelerate progress. This is aided by an ability to test approaches quickly and iterate solutions based on the results. If certain sub-problems inherently take a long time to solve (e.g. requiring lengthy real-time training in the physical world), this bounds the speed with which the overall project can be completed. Conversely, approaches that provide prompt feedback allow faster exploration of the search space of possible solutions—and thus, more effective application of resources.

### 3.1.6 *Behavior of key actors*

When AI safety and governance researchers attempt to assess how close the field is to the AGI runway, they will never have perfect information. Private-sector labs will keep proprietary details of their research closely guarded, and government-funded research may be classified. But the decisions and policies of those actors may provide indirect insight about the hidden information available to them.[11] For example, if scientists at top labs concede that they wouldn't know how to usefully spend a vastly larger budget, this is evidence that they don't see the surface area expanding. On the other hand, if a lab suddenly starts seeking and attracting vastly greater funding, this likely signals that they have discovered ways to productively apply those funds—which would imply expanding surface area.

Likewise, if a government abruptly starts changing its policies around AI research in a more nationalistic direction, or pouring opaque funding into new private-sector partnerships, this may be evidence that it perceives the AGI runway to have been reached.

Any tech company, investment firm, or state actor might be mistaken in its runway judgment. But to the extent those actors have established credibility and access to non-public information, unusual behavior changes should prompt the wider AGI community to ask "What do they know that we don't?"

## 4  CONCLUSIONS

In this paper, we have argued that while the 'surface area' of the AGI problem may currently be too small for states to apply massive resources, it is plausible that their incentives will change sharply once basic research sufficiently expands this surface area. Due to the scientific and geopolitical implications of such a shift, there is need for better metrics for assessing how close AGI research is to being 'sprintable.' As such, we have highlighted six broad elements of a roadmap that may approximate how the problem's surface area is expanding. However, these are admittedly still rough, and we therefore propose further research to formalize and refine such metrics, as well as to identify other criteria and desiderata. Such work, we hope, could contribute to an actionable 'roadmap to the roadmap.'

## ACKNOWLEDGEMENTS

We would like to thank Niall Ferguson, Eyck Freymann, Abie Katz, Jess Whittlestone, and one additional researcher who preferred not to be named for helpful conversations, as well as two anonymous reviewers for valuable comments and feedback.

## REFERENCES

[1] O. Afanasjeva, J. Feyereisl, M. Havrda, M. Holec, S. Ó hÉigeartaigh, M. Poliak, 'Avoiding the precipice: race avoidance in the development of artificial general intelligence', Medium, (2017).
[2] M.S. Armstrong, N. Bostrom, C. Shulman, 'Racing to the precipice: a model of artificial intelligence development', *AI & Society*, 31, 201–206, (2016).
[3] S.D. Baum, 'A survey of artificial general intelligence projects for ethics, risk, and policy', Global Catastrophic Risk Institute, 2017.
[4] N. Bostrom, *Superintelligence: Paths, Dangers, Strategies*, Oxford University Press, Oxford, U.K., 2014.
[5] T.B. Brown, B. Mann, N. Ryder, M. Subbiah, J. Kaplan, P. Dhariwal, A. Neelakantan, P. Shyam, G. Sastry, A. Askell, S.

---

[11] This is the same principle guiding our speculation in section 1.5 that scenario 2 (that we're already on the AGI runway) is relatively unlikely.

Agarwal, A. Herbert-Voss, G. Krueger, T. Henighan, R. Child, A. Ramesh, D.M. Ziegler, J. Wu, C. Winter, C. Hesse, M. Chen, E. Sigler, M. Litwin, S. Gray, B. Chess, J. Clark, C. Berner, S. McCandlish, A. Radford, I. Sutskever, D. Amodei, 'Language models are few-shot learners', ArXiv200514165 [cs], (2020).

[6] D. Castelvecchi, 'Hunt for gravitational waves to resume after massive upgrade', *Nature News & Comment*, 525, 301, (2015).

[7] CERN, 'Facts and figures about the LHC', (n.d.).

[8] China's State Council, 'A next generation artificial intelligence development plan', New America Cybersecurity Initiative, 2017.

[9] P. Cihon, M.M. Maas, L. Kemp, 'Should artificial intelligence governance be centralised?: design lessons from history', in: *Proceedings of the AAAI/ACM Conference on AI, Ethics, and Society*, ACM, 228–234, (2020).

[10] R. Crootof, 'Jurisprudential space junk: treaties and new technologies', in: C. Giorgetti, N. Klein (eds.), *Resolving Conflicts in the Law*, Brill/Nijhoff, Leiden, Netherlands/Boston MA, 106–129, 2019.

[11] A. Dafoe, 'AI governance: a research agenda', Center for the Governance of AI, Future of Humanity Institute, Oxford University, 2018.

[12] A. Demski, 'Alignment research field guide', AI Alignment Forum, 2019.

[13] A. Duettmann, O. Afanasjeva, S. Armstrong, R. Braley, J. Cussins, J. Ding, P. Eckersley, M. Guan, A. Vance, R. Yampolskiy, 'Artificial general intelligence: coordination & great powers', Foresight Institute, 2018.

[14] E. Duffin, 'Top 20 R&D spenders 2018', Statista, (2020).

[15] A.H. Eden, E. Steinhart, D. Pearce, J.H. Moor, 'Singularity hypotheses: an overview', in: A.H. Eden, J.H. Moor, J.H. Søraker, E. Steinhart (eds.), *Singularity Hypotheses*, Springer Berlin Heidelberg, 1–12, 2012.

[16] ESA, 'How much does it cost?', European Space Agency, (2010).

[17] T. Everitt, G. Lea, M. Hutter, 'AGI safety literature review', ArXiv180501109 [cs], (2018).

[18] H. Fountain, 'A dream of clean energy at a very high price', *New York Times*, (2017).

[19] E. Geist, A.J. Lohn, 'How might artificial intelligence affect the risk of nuclear war?', RAND, 2018.

[20] J. Gertler, 'F-35 Joint Strike Fighter (JSF) program', Congressional Research Service, 2020.

[21] B. Goertzel, C. Pennachin, eds., *Artificial General Intelligence*, Springer Berlin Heidelberg, 2007.

[22] I.J. Good, 'Speculations concerning the first ultraintelligent machine,' *Advances in Computers*, 6, 31–88, (1964).

[23] K. Grace, 'Leó Szilárd and the danger of nuclear weapons: a case study in risk mitigation', Machine Intelligence Research Institute, 2015.

[24] K. Grace, J. Salvatier, A. Dafoe, B. Zhang, O. Evans, 'When will AI exceed human performance? Evidence from AI experts', *Journal of Artificial Intelligence Research*, 62, 729–754, (2018).

[25] R. Gruetzemacher, J. Whittlestone, 'Defining and unpacking transformative AI', ArXiv191200747 [cs], (2019).

[26] M. Halina, J. Martin, 'Five ways AI is not like the Manhattan Project (and one way it is)', 3 Quarks Daily, (2019).

[27] T.A. Han, L.M. Pereira, T. Lenaerts, 'Modelling and influencing the AI bidding war: a research agenda', in: *Proceedings of the 2019 AAAI/ACM Conference on AI, Ethics, and Society*, ACM, 5–11, 2019.

[28] J. Haner, D. Garcia, 'The artificial intelligence arms race: trends and world leaders in autonomous weapons development', *Global Policy*, 10, 331–337, (2019).

[29] K. Hao, 'The messy, secretive reality behind OpenAI's bid to save the world', *MIT Technology Review*, (2020).

[30] D. Hawkins, 'Manhattan District history - Project Y - the Los Alamos Project', Los Alamos Scientific Laboratory of the University of California, 1946.

[31] G. Hinton, Geoffrey Hinton (@geoffreyhinton): "Extrapolating the spectacular performance of GPT3 into the future suggests that the answer to life, the universe and everything is just 4.398 trillion parameters." Twitter, (2020).

[32] A. Imbrie, E.B. Kania, 'AI safety, security, and stability among great powers: options, challenges, and lessons learned for pragmatic engagement', Center for Security and Emerging Technology, 2019.

[33] D.J. Jilk, 'Conceptual-linguistic superintelligence', *Informatica*, 41, 429–439, (2017).

[34] E. Kania, 'The pursuit of AI is more than an arms race', Defense One, (2018).

[35] J. Kaplan, S. McCandlish, T. Henighan, T.B. Brown, B. Chess, R. Child, S. Gray, A. Radford, J. Wu, D. Amodei, 'Scaling laws for neural language models', ArXiv200108361 [cs, Stat], (2020).

[36] H. Karnofsky, 'Some background on our views regarding advanced artificial intelligence', Open Philanthropy Project, 2016.

[37] M.T. Klare, 'AI arms race gains speed', *Arms Control Today*, 49, (2019).

[38] A. Knapp, 'How much does it cost to find a Higgs boson?', *Forbes*, (2012).

[39] R. Kurzweil, *The Singularity is Near*, Viking Press, 2005.

[40] R. Loosemore, B. Goertzel, 'Why an intelligence explosion is probable', in: A.H. Eden, J.H. Moor, J.H. Søraker, E. Steinhart (eds.), *Singularity Hypotheses*, Springer Berlin Heidelberg, 2012: pp. 83–98.

[41] M.M. Maas, 'Innovation-proof governance for military AI? how I learned to stop worrying and love the bot', *Journal of International Humanitarian Legal Studies*, 10, 129–157, (2019).

[42] M.M. Maas, 'How viable is international arms control for military artificial intelligence? Three lessons from nuclear weapons', *Contemporary Security Policy*, 40, 285–311, (2019).

[43] G. Marcus, 'Deep learning: a critical appraisal', ArXiv180100631 [cs, Stat], (2018).

[44] G. Marcus, 'Innateness, AlphaZero, and artificial intelligence', ArXiv180105667 [cs], (2018).

[45] J. McCarthy, 'From here to human-level AI', in: *Proceedings of the Fifth International Conference on Principles of Knowledge Representation and Reasoning*, 640–646, 1996.

[46] K. Mizokami, 'A new weapons rack just increased the F-35's missiles by 50 percent', *Popular Mechanics*, (2019).

[47] R. Mola, 'Super helmet', *Air & Space Magazine*, (2017).

[48] V.C. Müller, N. Bostrom, 'Future progress in artificial intelligence: a survey of expert opinion', in: Müller, Vincent C. (Ed.), *Fundamental Issues in Artificial Intelligence*, Synthese Library, Berlin, 2016.

[49] National Human Genome Research Institute, 'Human Genome Project FAQ', Genome.Gov., 2020.

[50] W. Naudé, N. Dimitri, 'The race for an artificial general intelligence: implications for public policy', *AI & Society*, 35, 367–379, (2019).

[51] S. Nellis, 'Microsoft to invest $1 billion in OpenAI', Reuters, (2019).

[52] Office of Science and Technology Policy, 'The National Artificial Intelligence Research and Development Strategic Plan', National Science and Technology Council, 2016.

[53] Office of Science and Technology Policy, 'American Artificial Intelligence Initiative: year one annual report', The White House, 2020.

[54] K. Payne, *Strategy, Evolution, and War: From Apes to Artificial Intelligence,* Georgetown University Press, Washington, DC, 2018.

[55] K. Payne, 'Artificial intelligence: a revolution in strategic affairs?', *Survival*, 60, 7–32, (2018).

[56] M. Robinson, 'The CERN community; a mechanism for effective global collaboration?', *Global Policy*, 10, 41–51, (2018).

[57] H.M. Roff, 'The frame problem: the AI "arms race" isn't one', *Bulletin of the Atomic Scientists*, 75, 95–98, (2019).

[58] A. Roland, P. Shiman, *Strategic Computing: DARPA and the Quest for Machine Intelligence, 1983-1993*, MIT Press, Cambridge, MA, 2002.

[59] M. Rosa, O. Afanasjeva, W. Millership, 'Report from the AI Race Avoidance Workshop', GoodAI and AI Roadmap Institute, Tokyo, 2017.

[60] A.A. Stein, 'Coordination and collaboration: regimes in an anarchic world', *International Organization*, 36, 299–324, (1982).

[61] D.D. Stine, 'The Manhattan Project, the Apollo program, and federal energy technology R&D programs: a comparative analysis', Congressional Research Service, 2009.

[62] R. Sutton, 'The bitter lesson', incompleteideas.net, 2019.
[63] A.M. Turing, 'Computing machinery and intelligence', *Mind*, 59, 433–460, (1950).
[64] E. Turner, Elliot Turner (@eturner303): "Reading the OpenAI GPT-3 paper. Impressive performance on many few-shot language tasks. The cost to train this 175 billion parameter language model appears to be staggering: Nearly $12 million dollars in compute based on public cloud GPU/TPU cost models (200x the price of GPT-2) https://t.co/5ztr4cMm3L" Twitter, (2020).
[65] United States Government Accountability Office, 'Defense acquisitions: assessments of selected weapon programs', United States Government Accountability Office, 2015.
[66] K. Wiggers, 'OpenAI launches an API to commercialize its research', VentureBeat, (2020).
[67] E. Yudkowsky, 'There's no fire alarm for artificial general intelligence', Machine Intelligence Research Institute, 2017.
[68] B. Zhang, A. Dafoe, 'Artificial intelligence: American attitudes and trends', Center for the Governance of AI, Future of Humanity Institute, University of Oxford, 2019.
[69] R. Zwetsloot, H. Toner, J. Ding, 'Beyond the AI arms race: America, China, and the dangers of zero-sum thinking', *Foreign Affairs*, (2018).